%% file: text.tex
\input macros.tex

\documentclass[preprint,aps]{revtex4}

\usepackage{graphics}		% Figures
\usepackage{amsmath,amssymb}	% AMS fonts and symbols
\usepackage{dcolumn}		% Align table columns on decimal point

\begin{document}

\title{A comparison of the static and dynamic properties of a semi-flexible 
polymer using lattice-Boltzmann and Brownian dynamics simulations}

\author{Anthony J. C. Ladd}
\author{Rahul Kekre}
\author{Jason E. Butler}
\affiliation {Department of Chemical Engineering, University of Florida, Gainesville, Florida 32611-6005, USA} 
\date{\today}

\preprint{Submitted to Phys. Rev. E}

\begin{abstract}
The aim of this paper is to compare results from lattice-Boltzmann and Brownian dynamics simulations of linear chain molecules. We have systematically varied the parameters that may affect the accuracy of the lattice-Boltzmann simulations, including grid resolution, temperature, polymer mass, and fluid viscosity. The effects of the periodic boundary conditions are minimized by an analytic correction for the different long-range interactions in periodic and unbounded systems. Lattice-Boltzmann results for the diffusion coefficient and Rouse mode relaxation times were found to be insensitive to temperature, which suggests that effects of hydrodynamic retardation are small. By increasing the resolution of the lattice-Boltzmann grid with respect to the polymer size, convergent results for the diffusion coefficient and relaxation times were obtained; these results agree with Brownian dynamics to within 1--2\%.
\end{abstract}
\maketitle

\section{Introduction}

The fluctuating lattice-Boltzmann (FLB) equation~\cite{Ladd1993} has been proposed as a basis for numerical simulations of polymer solutions~\cite{Ahlrichs1999}, using a frictional coupling between the polymer and the surrounding fluid. Single molecule dynamics obtained with this algorithm compared favorably with coarse-grained molecular dynamics simulations using explicit solvent~\cite{Dunweg1993}. More recently, the same idea has been investigated for confined polymers~\cite{Usta2005}, as a simpler and possibly more efficient alternative to Brownian dynamics (BD)~\cite{Jendrejack2002}.  Both methods have been applied to problems of polymer migration in shear and pressure-driven flows~\cite{Jendrejack2002,Jendrejack2004,Usta2006} and showed similar trends with increasing shear rate, but different choices of polymer model precluded a quantitative comparison. Here we initiate a systematic comparison of FLB and BD methods, beginning with the properties of an isolated chain; subsequently we will extend the investigation to confined polymers in shear and pressure-driven flows. This work complements a recent study by Pham et al.~\cite{Pham2009}.
 
Inertial effects are neglected in Brownian dynamics, resulting in instantaneous propagation of momentum. Although the solvent degrees of freedom are thereby eliminated, the interactions between the beads are long-range, which leads to an ${\cal O}(N^3)$ scaling of the computational cost for a polymer consisting of $N$ segments. In the FLB method, all interactions are local and the computational cost scales linearly with the volume. However, the lattice-Boltzmann method introduces an extra, inertial time scale, during which the hydrodynamic interactions propagate throughout the fluid by viscous momentum diffusion. Surprisingly this has little effect on the time step; FLB simulations use comparable time steps to BD, as will be seen in Sec.~\ref{static}. Hydrodynamic retardation, sometimes thought to be a potential source of error in lattice-Boltzmann simulations of Stokes flow~\cite{Chen2007,Pham2009}, is in fact easily managed. Nevertheless, the additional degrees of freedom of an explicit solvent model adds considerably to the computational cost. Generally, dilute systems in unconfined geometries favor BD, while more concentrated solutions in confined geometries favor the FLB method~\cite{Pham2009}.

Quantitative comparisons require an identical micro-mechanical model of the polymer. However, the BD simulations are for an isolated chain, while the FLB simulations use periodic boundary conditions. It is therefore necessary to correct for the differences in the long-range flow fields in periodic and unbounded systems; in particular, the diffusion coefficient in a periodic system has a correction proportional to $L^{-1}$, where $L$ is the length of the periodic unit cell. However, prior research~\cite{Ladd1990a,Usta2005} has shown that these corrections can be calculated quantitatively, based on the hydrodynamic theory for a periodic unit cell~\cite{Hasimoto1959}. Corrections to the configurational properties and Rouse relaxation times are much smaller, ${\cal O} (L^{-3})$; when the box length is $5-10$ times the polymer size, deviations from the infinite system are negligible. The diffusion coefficients and Rouse relaxation times were found to depend on the degree of discretization of the lattice-Boltzmann fluid. Our results show numerical convergence with increasing grid resolution, and the converged results agree with Brownian dynamics within $1-2\%$. 

\section{Polymer model and simulation methods}\label{model}

The polymer model consists of $N+1$ beads connected by FENE springs between neighboring beads,
\begin{equation}
    \Phi_{S} = \sum_{i = 0}^N \phi_S(|\vec{r}_{i+1} - \vec{r}_i|),  ~~~ \phi_S(r) =  -\frac{1}{2} \kappa r_{0}^2 \ln \left( 1 - \frac{r^2}{r_{0}^2}\right),
\label{eq:fene}
\end{equation}
where $\kappa$ is the spring constant, $r_{0}$ is the maximum extension of the spring, and $\vec{r}_{i}$ is the position vector of the $i^{th}$ bead. The simulations use a value of $r_{0} = 5.48b$, where $b = \sqrt{T/ \kappa }$ and $T$ is the thermal energy corresponding to an absolute temperature $T/k_B$. The root-mean-square bond length in an ideal chain, $\left<r^2\right>^{1/2} = 1.60b$, follows from the potential in Eq.~\eqref{eq:fene};
\begin{equation}\label{eq:r2}
\left<r^2\right> = \frac{3T}{\kappa}\left(\frac{r_0^2}{r_0^2+5b^2}\right).
\end{equation} 
In addition to the FENE potential, there is an excluded volume interaction between the beads,
\begin{equation}
    \Phi_{EV} = \sum_{i>j} \phi_{EV}(|r_{i} - r_{j}|),~~~ \phi_{EV}(r)=  \epsilon \exp(-\beta r^2),
\label{eq:ev}
\end{equation}
with $\beta = 1.50 b^{-2}$ and  $\epsilon = 2.71 T$. The potential energy $\Phi = \Phi_S + \Phi_{EV}$ approximates a DNA molecule with $\sim 10$ Kuhn segments per spring~\cite{Jendrejack2000,Jendrejack2002}. The identical polymer model, described by Eqs.~\eqref{eq:fene} and~\eqref{eq:ev}, was used for both BD and FLB simulations. The beads are coupled to the fluid with a Stokes friction coefficient $\xi = 6 \pi \eta a$, where $\eta$ is the fluid viscosity and the hydrodynamic radius of the beads $a = 0.362b$. 

\subsection{Brownian dynamics}
\label{BD}

Brownian dynamics neglects inertia, and the state of the polymer is therefore completely specified by the positions of the $N+1$ beads, $\Vr_i$. Hydrodynamic interactions (HI) between the beads are introduced in a pairwise-additive approximation through the mobility matrix $\Vmu_{ij}$, which connects the mean (or drift) velocity of bead $i$ to the force on bead $j$,
\begin{equation}
{\bar \Vv}_i = \sum_{j=0}^N \Vmu_{ij} \cdot \VF_j.
\end{equation}
The conservative force, $\VF_j = -\Dx_{{\Vr}_j} \Phi$, is derived from the potential energy of the polymer, Eqs.~\eqref{eq:fene} and \eqref{eq:ev}. We use the Rotne-Prager regularization of the mobility matrix of point particles~\cite{Rotne1969,Yamakawa1970}, which approximates the Stokes-flow result in the limit $r_{ij} \gg a$, but ensures that the mobility matrix remains positive definite for all $r_{ij}$:
\begin{equation} \label{eq:rpy}
\Vmu_{ij} = \xi^{-1}
\begin{cases} 
\begin{array}{cc}
C_1 \VI + C_2 \dfrac{{\Vr}_{ij}\otimes{\Vr}_{ij}}{r_{ij}^2} & r_{ij} > 2a \\
C_1^\prime \VI  + C_2^\prime \dfrac{{\Vr}_{ij}\otimes{\Vr}_{ij}}{r_{ij}^2} , & r_{ij} \leq 2a
\end{array}
\end{cases}
\end{equation}
where
\begin{eqnarray}
\begin{array}{ll}
C_1 = \dfrac{3}{4} \dfrac{a}{r_{ij}} + \dfrac{1}{2} \dfrac{a^3}{r_{ij}^3} , & C_2 =   \dfrac{3}{4} \dfrac{a}{r_{ij}}  - \dfrac{3}{2} \dfrac{a^3}{r_{ij}^3}, \\
C_1^\prime = 1 - \dfrac{9}{32} \dfrac{r_{ij}}{a}, & C_2^\prime =  \dfrac{3}{32} \dfrac{r_{ij}}{a}.
\end{array}
\end{eqnarray}
When $i = j$, the self mobility, $\Vmu_{ii} = \xi^{-1} \VI$, is the mobility of an isolated sphere.

The Rotne-Prager mobility is divergence free, $\sum_{i=0}^N \Dx_{{\Vr}_i} \cdot \Vmu_{ij} = 0$, and the first order Ermak and McCammon algorithm~\cite{Ermak1978} for stochastic integration reduces to an explicit Euler integration scheme,
\begin{equation}
\Vr_i(t+\dt) =\Vr_i(t) + {\bar \Vv}_i \dt + \Delta \Vw_i ,
\label{eq:langvin}
\end{equation}
where $\dt$ is the time step and $\Delta \Vw_i$ is a random displacement with zero mean and covariance
\begin{equation}
<\Delta \Vw_i \otimes \Delta \Vw_j > = 2 T \Vmu_{ij} \dt.
\label{eq:fdt}
\end{equation}
We used a Cholesky decomposition of the grand mobility matrix,
\begin{equation}\left[
\begin{array}{cccc}
 \Vmu_{00} &  \Vmu_{01} & \ldots & \Vmu_{0,N} \\
 \Vmu_{10} &  \Vmu_{11} & \ldots & \Vmu_{1,N} \\
\vdots &  \ldots & \ldots & \vdots \\
 \Vmu_{N,0} &  \Vmu_{N,1} & \ldots & \Vmu_{N,N}
\end{array}
\right],
\end{equation} 
to calculate the random displacements, which is an $O(N^3)$ computation. However, the Cholesky decomposition is a small fraction of the total computational cost for the short chains ($N = 10$, $20$) used in this work, and eliminates any possible errors associated with the Chebyshev polynomial approximation~\cite{Fixman1986}, which scales more favorably as $O(N^{2.25})$. 

\subsection{Lattice Boltzmann}
\label{FLB}
The fluctuating lattice-Boltzmann model~\cite{Ladd1993a,Ladd1994} has been used to simulate the dynamics of dilute polymer solutions in periodic~\cite{Ahlrichs1999,Ahlrichs2001} and confined geometries~\cite{Usta2005,Usta2006,Usta2007}. In the original formulation of the FLB model~\cite{Ladd1993a,Ladd1994}, the viscous stress tensor was assumed to fluctuate around the local Navier-Stokes stress, but this model fails to satisfy the fluctuation-dissipation theorem at small scales~\cite{Adhikari2005} unless thermal fluctuations in the non-hydrodynamic modes are included as well. Here we summarize the improved method, following the recent statistical-mechanical formulation of the FLB equation~\cite{Dunweg2007}. Including thermal fluctuations in the non-hydrodynamic modes leads to small, but noticeable, improvements in the equipartition of energy between the fluid and polymer degrees of freedom (Sec.~\ref{static}).

In the lattice-Boltzmann (LB) model, the fluid degrees of freedom are represented by a discretized one-particle velocity distribution function $n_i(\Vr,t)$, which describes the mass density of particles with velocity $\Vc_i$ at the position ${\Vr}$ and time $t$.  The hydrodynamic fields, mass density $\rho$, and momentum density $\Vj = \rho \Vu$, are moments of this velocity distribution,
\begin{eqnarray}\label{eq.Moments}
\begin{array}{cc}
\rho = \sum_i n_i, & \Vj = \sum_i n_i\Vc_i.
\end{array}
\end{eqnarray}
The time evolution of $n_i(\Vr, t)$ is described by a discrete analogue of the Boltzmann equation~\cite{Frisch1987},
\begin{equation}\label{eq.Tevol}
n_i(\Vr + \Vc_i \dt,t + \dt) = n_i(\Vr,t) + \Delta_i\left[\Vn (\Vr,t)\right];
\end{equation}
here $\Delta_i$ is the change in $n_i$ due to instantaneous collisions at the lattice nodes and $\dt$ is the time step.

The D3Q19 model~\cite{Qian1992} was used, which includes rest particles and 18 velocities corresponding to the [100] and [110] directions of a simple cubic lattice. The population density associated with each velocity has a weight $a^{c_i}$ that describes the fraction of particles with velocity $\Vc_i$ in a fluid at rest:
\begin{equation}\label{eq.Weights}
a^0 = \frac{1}{3}, \hspace{1em} a^1 = \frac{1}{18}, \hspace{1em} a^{\sqrt 2} = \frac{1}{36}.
\end{equation}
The deterministic collision operator is typically linearized about the low-velocity equilibrium distribution, $n_i^{eq}$,
\begin{equation} \label{eq:nieq}
n_i^{eq} \left(\rho, \Vu \right) = a^{c_i} \rho \left( 1 + \frac{\Vu \cdot \Vc_i}{c_s^2} + \frac{2 \Vu \Vu : (\Vc_i \Vc_i - c_s^2 \Vone)}{c_s^4} \right),
\end{equation}
where the speed of sound $c_s =  3^{-1/2} \dx / \dt$ and $\dx$ is the lattice spacing. The non-equilibrium distribution, $n_i^{neq} = n_i - n_i^{eq}$,  can then be expanded in moments~\cite{dHumieres1992,dHumieres2002}, 
\begin{equation}\label{eq:ni-mk}
m_k = \sum_i n_i^{neq} e_{ki},
\end{equation} 
using tensorial polynomials of the lattice vectors, $e_{ki}$ as a basis. We use a different basis from Refs.~\cite{dHumieres1992,dHumieres2002}, such that the back transformation includes the weights $a^{c_i}$ \cite{Chun2007},
\begin{equation}
n_i^{neq} = a^{c_i}\sum_k w_k^{-1}e_{ki}m_k,
\end{equation}
where $w_k = \sum_i a^{c_i} e_{ki}^2$ is the normalizing factor for $\Ve_k$. During the collision process, the moments $m_k$ relax towards equilibrium (zero),
\begin{equation}\label{eq:mk*}
m_k^\star = \gamma_k m_k,
\end{equation}
where the relaxation parameter is bounded by $\modulo{\gamma_k} < 1$. In these simulations we used a two-parameter collision operator, with different eigenvalues for the modes with odd ($\gamma_o$) and even  ($\gamma_e$) powers of $\Vc_i$ \cite{Ginzburg2003}. The shear viscosity is related to $\gamma_e$,
\begin{equation}\label{eq:vis}
\eta = \frac{\rho c_s^2 h}{2}\left(\frac{1+\gamma_e}{1-\gamma_e}\right),
\end{equation}
and
\begin{equation}\label{eq:oddeven}
\gamma_o = -\frac{7 \gamma_e + 1}{\gamma_e + 7};
\end{equation} 
this relation makes the location of a planar solid boundary independent of viscosity~\cite{Ginzburg2003,Chun2007}, although that property is not essential in the present context.

The key difference between the FLB and LB models is in the collision operator. The FLB collision operator contains random excitations of the non-conserved moments, \cf Eq.~\eqref{eq:mk*}~\cite{Dunweg2009},
\begin{equation}\label{eq:mk*f}
m_k^\star = \gamma_k m_k + \sqrt{\frac{\rho m_p w_k (1-\gamma_k^2)}{\dx^3}} \phi_k,
\end{equation}
where $\phi_k$ is a random variable with zero mean and unit variance. It is important to use a bounded distribution of random numbers~\cite{Dunweg2009}, or large changes in $m_k$ will occasionally occur, leading to negative values of $n_i$. The amplitude of the random forcing is determined from the fluctuation-dissipation relation~\cite{Ladd1994} and is controlled by the mass of an LB ``particle'', $m_p$~\cite{Dunweg2007}. Thermodynamic consistency requires that $m_p$ is related to the effective temperature of the fluctuating fluid~\cite{Dunweg2007,Dunweg2009},
\begin{equation} \label{eq:m_p}
m_p c_s^2 = T.
\end{equation}
In this work the random forcing is applied to all the non-conserved modes~\cite{Adhikari2005}, not just the stress~\cite{Ladd1993}. It has been shown theoretically~\cite{Dunweg2007} that the excitation of the non-hydrodynamic modes is essential to satisfy the fluctuation-dissipation relation at all scales, although at long wavelengths only the excitations in stress are important.

\subsection{Polymer-fluid coupling}\label{sec:coupling}

The polymer is coupled to the LB fluid by frictional forces between the beads and the fluid. The equations of motion for the $i^{th}$ monomer can be written in inertial form as,
\begin{equation}\label{eq:EOM}
 \frac{d\Vr_i}{dt} = \Vv_i, ~~~ m \frac{d\Vv_i}{dt} = \VF_i -\xi_0 \left[\Vv_i(t) - \Vu(\Vr_i,t)\right] + \VR_i(t).
\end{equation}
where the hydrodynamic force includes a frictional drag, based on the difference in velocity between the bead and the surrounding fluid, and a random force, $\VR_i$, to balance the additional dissipation~\cite{Ahlrichs1999}. Since the fluid satisfies its own fluctuation-dissipation relation, $\VR_i$ has a local covariance matrix
\begin{equation}\label{eq:rforce}
    \left<\VR_i(t) \VR_j(t^\prime)\right> = 2T \xi_0 \delta(t-t^\prime) \delta_{ij} \mathbf{I} .
\end{equation}
Hydrodynamic interactions between the beads are transmitted through the fluid via correlated fluctuations in the velocity field, which develop over the inertial time scale, $\rho r^2/\eta$, where $r$ is the separation between beads.  The large time-scale separation between the dynamics of the polymer and the individual monomers allows time for the hydrodynamic interactions to reach a quasi-steady state, without imposing this condition at each and every time step. We will show numerically that both inertial (FLB) and diffusive (BD) simulations can use similar time steps, of the order of the monomer diffusion time (see also Ref.~\cite{Usta2005}).

Since the monomers move continuously over the grid, the fluid velocities, $\Vu_n$, are interpolated from neighboring grid points to the bead location $\Vr_i$,
\begin{equation}\label{eq:interp}
    \Vu(\Vr_i,t) = \sum_n \Delta(\Vr_i-\Vr_n) \Vu_n.
\end{equation}
The interpolating function $\Delta(r_x, r_y, r_z)$ is taken as a product of one-dimensional functions~\cite{Pes02}
\begin{equation}\label{eq:Delta}
\Delta(x,y,z) = \phi\left(\frac{x}{\dx}\right)
        \phi\left(\frac{y}{\dx}\right) \phi\left(\frac{z}{\dx}\right) ,
\end{equation}
where $\dx$ is the grid spacing in the lattice-Boltzmann simulations. Typically the weights $\phi(u)$ are determined by linear (two-point) interpolation,
\begin{equation}\label{eq:2point}
\phi_2(u) = \left\{
\begin{array}{lcl}
1 - \modulo{u} & \hspace{3em} & \modulo{u} \le 1 , \\ 
0           & \hspace{3em} & \modulo{u} \ge 1 ,
\end{array} \right.
\end{equation}
but a more precise interpolation is possible using three or four points in each coordinate direction. Numerical tests of the different interpolations can be found in Ref.~\cite{Dunweg2009}. We will present some results with three-point interpolation,
\begin{equation}\label{eq:3point}
\phi_3(u) = \left\{
\begin{array}{lcl}
\frac{1}{3} \left(1 + \sqrt{1-3u^2}\right)
&\hspace{3em}&
0 \le \modulo{u} \le \frac{1}{2}\\
\frac{1}{6} \left(5-3\modulo{u} - \sqrt{-2+6\modulo{u}-3u^2}\right)
&\hspace{3em}&
\frac{1}{2} \le \modulo{u} \le \frac{3}{2}\\
0
&\hspace{3em}&
\frac{3}{2} \le \modulo{u},
\end{array} \right.
\end{equation}
but most of the results use $2$-point (linear) interpolation. To conserve momentum, the accumulated force exerted by the bead on the fluid is distributed to the surrounding nodes with the same weight function~\cite{Ahlrichs1999,Dunweg2009}. 

The input friction $\xi_0 = 6 \pi \eta a_0$ is not the same as the effective friction $\xi = 6 \pi \eta a$, as measured by the drag force on the bead or by its diffusion coefficient. This is because the force added to the fluid renormalizes the input friction,
\begin{equation} \label{eq:aeff}
\frac{1}{\xi} = \frac{1}{\xi_0} + \frac{1}{6\pi \eta \dx g},
\end{equation}
where $g$ is a numerical factor~\cite{Dunweg2009} that is independent of the fluid viscosity but depends on the interpolation function. From the diffusion of individual monomers we have determined values of $g = 1.3$ for  linear (two-point) interpolation and $g = 1.0$ for the three-point interpolation. The FLB results in this paper are matched to BD simulations with the same effective radius, $a$.

The coupled equations of motion for the particles and fluid are solved by operator splitting~\cite{Dunweg2009}; typically, the thermodynamic forces are integrated with a smaller time step than the hydrodynamic forces to maintain stability. The LB time interval $\dt$ is decomposed into $M$ steps of length $h = \dt/M$, where $M$ is chosen to be sufficiently large that  the conservative forces are integrated accurately; typically $M \sim 10$ in our simulations. The algorithm used in this work is as follows:
\begin{enumerate}
\item At the beginning of the LB step, determine the fluid velocities at the grid points.
\item Update the polymer positions and velocities over $M$ sub-cycles. For each subcycle a modified Verlet algorithm is used to update the positions and velocities of the beads:
\begin{enumerate}
\item  First stream the particle positions and velocities for half a time step,
\begin{eqnarray}\label{eq:p1}
&&\Vr_i^{(1)} = \Vr_i(t) + \dfrac{h}{2}\Vv_i(t), \\
\label{eq:v1}
&&\Vv_i^{(1)}= \Vv_i(t) +  \dfrac{h}{2m}\VF_i^{(1)},
\end{eqnarray}
where the conservative force $\VF_i^{1} = -\Dx_{\Vr_i^{(1)}}\Phi$ is evaluated from the coordinates at the half time step, $\Vr_i^{(1)}$.
\item Use the updated positions, $\Vr_i^{(1)}$, to interpolate the fluid velocity to the bead locations, Eq.~\eqref{eq:interp}.
\item Update the bead velocities for a full step $h$, using a midpoint approximation to the frictional drag force Eq.~\eqref{eq:EOM},
\begin{equation}\label{eq:v}
 m \frac{\Vv_i^{(2)}-\Vv_i^{(1)}}{h} = -\xi_0 \left( \frac{\Vv_i^{(2)} + \Vv_i^{(1)}}{2} - \Vu(\Vr_i,t)\right) + \sqrt{\frac{2 T \xi_0}{h}}\Vphi_i,
\end{equation}
where $\Vphi_i$ is a vector of bounded random numbers with zero mean and unit variance.
\item Redistribute the momentum transferred by particle-fluid coupling
\begin{equation}\label{eq:dp}
\Delta \Vp_i = \frac{-\xi_0 h \left(\Vv_i^{(1)} -\Vu(\Vr_i,t)\right) + \sqrt{2 T \xi_0 h}\Vphi_i}{1 + \xi_0 h/2m},
\end{equation}
back to the fluid.
\item Stream the particle positions and velocities for the second half step,
\begin{eqnarray}\label{eq:v2}
&&\Vv_i(t+h)= \Vv_i^{(2)} +  \dfrac{h}{2m}\VF_i^{(1)}, \\
\label{eq:p2}
&&\Vr_i(t+h) = \Vr_i^{(1)} + \dfrac{h}{2}\Vv_i(t+h), \\
\end{eqnarray}
\end{enumerate}
The exact sequence of updates is important to preserve the second-order accuracy of the operator-splitting method. The algorithm reduces to the Verlet scheme when $\xi_0 \rightarrow 0$.
\item Update LB populations to account for momentum transfer from particle-fluid coupling.
\item Update FLB algorithm for one time step $\dt$.
\end{enumerate}

There are a number of nearly equivalent ways to break down the coupled dynamics of the particle-fluid system; the algorithm described above is the most accurate of the variations we have investigated, although the differences in long-time properties (conformation, diffusion, Rouse relaxation times) are generally small. The midpoint method is preferable to a first-order update, either explicit or implicit, since neither of these lead to exact thermalization of the kinetic energy of the particles. For force-free particles it is straightforward to show that~\cite{Usta2005}
\begin{equation}
m\left<v_i^2\right> = \frac{3T}{1 \pm \xi_0 h/m},
\end{equation}
with the plus sign following from the implicit update and the minus sign from the explicit update. By contrast, the midpoint method gives $m\left<v_i^2\right> = 3T$ exactly. There is a choice as to whether the momentum transferred to the fluid, Eq.~\eqref{eq:dp}, affects the interpolated fluid velocity after each sub step ($h$) or only after each LB step ($\dt$). Numerical results show that the polymer temperature is closer to the fluid temperature (within $0.3\%$) if the interpolated velocity is updated every sub step ($h$). When the velocity is only updated at the LB steps (every $\dt$), the polymer temperature differs from the fluid temperature by about $3\%$. The results presented in Sec.~\ref{sec:results} have the interpolated fluid velocity updated every $h$.

\section{Results}\label{sec:results}

The purpose of these simulations was to make precise comparisons of BD and FLB results for an identical polymer model. We have compared static properties (radius of gyration and end-to-end distance) and dynamic properties (diffusion coefficient and Rouse relaxation times). The effects of time step have been investigated, and, in the case of the FLB simulations, the effects of grid resolution, temperature (fluctuation amplitude), fluid viscosity, and bead mass as well. To obtain statistically precise data, every FLB data point was calculated from an ensemble average over 160 different initial conditions. Each sample was equilibrated for a time of approximately $500 t_0$ and data was collected for a further $5000 t_0$, for a total of $8 \times 10^5 t_0$. The time unit $t_0 = \xi/\kappa$, and the Zimm time $t_Z = 6\pi \xi R_g^3/T = 56.7t_0$. The Brownian dynamics simulations of diffusion and Rouse relaxation times were run for $1.6 \times 10^6 t_0$ and $8 \times 10^6 t_0$ respectively.

\subsection{Static properties}\label{static}

\begin{figure}
\scalebox{0.5}{\includegraphics{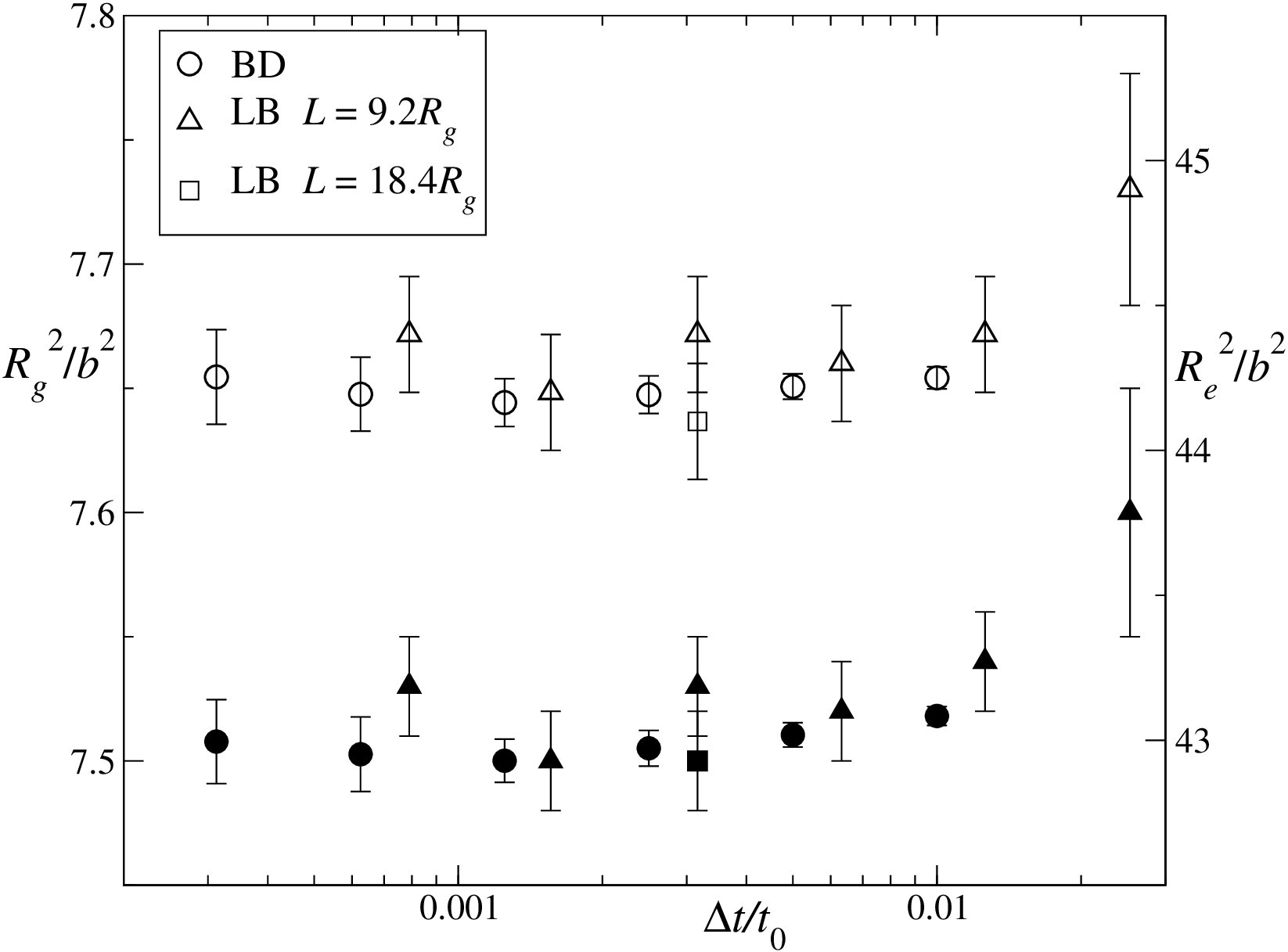}}
\caption{Conformational properties, $R_{g}^{2}/b^2$ (closed symbols) and $R_{e}^{2}/b^2$ (open symbols), versus the dimensionless time step $\dt / t_{0}$. The FLB data used a grid resolution $\dx = 1.29b$; conformational properties with other grid resolutions are statistically indistinguishable (see Table~\ref{tab:props}).}
\label{fig:conf}
\end{figure}

The mean square radius of gyration, $R_g^2$, 
\begin{equation}
 \left \langle R_g^2 \right \rangle = \frac{1}{2( N+1)^2} \sum_{ij} \left \langle r_{ij}^2 \right \rangle, 
\label{eq:rg}
\end{equation}
and end-to-end vector, $R_e^2$,
\begin{equation}
\left \langle R_e^2 \right \rangle = \left \langle (\Vr_N - \Vr_0)^2 \right \rangle,
\label{eq:re}
\end{equation}
of the polymer are compared in Fig.~\ref{fig:conf}. The conformational properties from lattice-Boltzmann are indistinguishable from Brownian dynamics within the statistical errors ($0.5\%$). Despite the extra inertial time scale, the FLB simulations use comparable timesteps to Brownian dynamics; neither method shows statistically significant deviations in $R_g$ and $R_e$ when the time step $\dt$ is less than $0.01 t_0$.

The accuracy of a BD simulation depends only on the time step, but results from FLB simulations may depend on a number of parameters: fluctuation level or temperature ($T$), length of the periodic unit cell ($L$), grid resolution ($\dx$), fluid viscosity ($\eta$), and particle mass ($m$). Results for a range of values of these parameters are summarized in Table~\ref{tab:props} using a FENE chain (Sec.~\ref{model}) of $10$ segments. In FLB simulations, the timescale $t_0 = \xi/\kappa = \xi b^2/T$ is controlled by the temperature, which sets the level of fluctuations in the fluid, Eqs.~\eqref{eq:mk*f}--\eqref{eq:m_p}, and particles, Eqs.~\eqref{eq:EOM}--\eqref{eq:rforce}. The temperature reflects the degree of coarse graining of the molecular fluid rather than the thermodynamic properties of the chain~\cite{Dunweg2007,Dunweg2009}; one dimensionless measure is the parameter $\alpha = \left<u_1^2\right>/c_s^2$ (see Table~\ref{tab:props}), which relates the fluid velocity fluctuations in a single LB cell to the sound speed. For the LB model to adequately represent an incompressible fluid ($|u|/c_s < 0.1$), $\alpha$ should be less than $0.01$. At the highest temperature shown in Table~\ref{tab:props}, $\alpha = 0.024$, the polymer is indeed slightly swollen. 

\begin{table}
\renewcommand*\arraystretch{0.6}
\begin{tabular}{|ccD{.}{.}{5}ccc|cc|ccc|}
\hline
$\dx/b$ && \multicolumn{1}{c}{$\alpha$} & $\dt /t_0$ &
$\Delta T/T$ & $Sc_0$ & $R_e^2/b^2$ & $R_g^2/b^2$ & $D/D_0$ & $Sc$ & $\tau_1/t_0$ \\ \hline
2.58 && 0.003 & 2.53(-2) & 0.003 & 26 & 44.3 $\pm$ 0.2 & 7.52 $\pm$ 0.02 & 0.188 & 127 & 19.0 \\
2.58 &$^{1}$& 0.003 &1.26(-2) & 0.003 & 53 & 44.2 $\pm$ 0.2 & 7.52 $\pm$ 0.02 & 0.187 & 254 & 19.2 \\
1.29 && 0.024 & 2.53(-2) & 0.051 & 7 & 44.9 $\pm$ 0.4 & 7.60 $\pm$ 0.05 & 0.203 & 32 & 18.0 \\ 
1.29 && 0.012 & 1.26(-2) & 0.002 & 13 & 44.4 $\pm$ 0.2 & 7.54 $\pm$ 0.02 & 0.206 & 64 & 17.4 \\
1.29 && 0.006 & 6.32(-3) & 0.002 & 26 & 44.3 $\pm$ 0.2 & 7.52 $\pm$ 0.02 & 0.205 & 128 & 17.1 \\
1.29 && 0.003 & 3.16(-3) & 0.002 & 53 & 44.4 $\pm$ 0.2 & 7.54 $\pm$ 0.02 & 0.206 & 256 & 17.3 \\
1.29 &$^1$& 0.003 & 3.16(-3) & 0.002 & 53 & 44.1 $\pm$ 0.2 & 7.50 $\pm$ 0.02 & 0.204 & 258 & 17.3 \\
1.29 &$^{2a}$& 0.003 & 6.32(-3) & 0.001 & 13 & 44.2 $\pm$ 0.2 & 7.51 $\pm$ 0.02 & 0.205 & 64 & 17.7 \\
1.29 &$^{2b}$& 0.003 & 1.58(-3) & 0.004 & 211 & 44.4 $\pm$ 0.2 & 7.53 $\pm$ 0.02 & 0.205 & 1030 & 17.3 \\
1.29 &$^{3a}$& 0.003 & 3.16(-3) & 0.002 & 53 & 44.2 $\pm$ 0.2 & 7.49 $\pm$ 0.02 & 0.205 & 257 & 17.1 \\
1.29 &$^{3b}$& 0.003 & 3.16(-3) & 0.001 & 53 & 44.2 $\pm$ 0.2 & 7.51 $\pm$ 0.02 & 0.203 & 260 & 17.3 \\
1.29 &$^{4}$& 0.003 & 3.16(-3) & 0.001 & 53 & 44.2 $\pm$ 0.2 & 7.51 $\pm$ 0.02 & 0.199 & 265 & 17.8 \\
1.29 &$^{5}$& 0.003 & 3.16(-3) & 0.011 & 53 & 43.2 $\pm$ 0.5 & 7.27 $\pm$ 0.04 & 0.196 & 269 & 18.5 \\
1.29 &$^{6}$& 0.003 & 3.16(-3) & 0.026 & 53 & 37.4 $\pm$ 0.1 & 6.28 $\pm$ 0.01 & 0.098 & 540 & 28.3 \\
1.29 && 0.0015 & 1.58(-3) & 0.002 & 106 & 44.2 $\pm$ 0.2 & 7.50 $\pm$ 0.02 & 0.205 & 514 & 17.5 \\
1.29 && 0.00075 &7.90(-4) & 0.002 & 211 & 44.4 $\pm$ 0.2 & 7.53 $\pm$ 0.02 & 0.205 & 1028 & 17.5 \\
0.86 && 0.003 & 9.37(-4) & 0.002 & 79 & 44.2 $\pm$ 0.2 & 7.50 $\pm$ 0.02 & 0.210 & 377 & 16.2 \\
0.86 &$^{4}$& 0.003 & 9.37(-4) & 0.002 & 79 & 44.2 $\pm$ 0.2 & 7.51 $\pm$ 0.02 & 0.208 & 380 &  \\
0.65 && 0.003 & 3.95(-4) & 0.002 & 106 & 44.1 $\pm$ 0.2 & 7.49 $\pm$ 0.02 & 0.210 & 502 & 16.1 \\ \hline
BD &$^7$&& 3.13(-4) &&& 44.2 $\pm$ 0.1 & 7.50 $\pm$ 0.01 & 0.208 & $\infty$ & 16.5 \\
BD &$^7$&& 6.25(-4) &&& 44.2 $\pm$ 0.1 & 7.50 $\pm$ 0.01 & 0.209 & $\infty$ & 16.4 \\
BD &$^7$&& 1.25(-3) &&& 44.3 $\pm$ 0.2& 7.51 $\pm$ 0.01 & 0.208 & $\infty$ & 16.5 \\ \hline
\end{tabular}

\caption{Static and dynamic properties of a polymer chain. Fluctuating LB simulations for a 10-segment chain are compared with Brownian dynamics. The resolution of the LB grid, $\dx$, can be compared with the RMS distance between the beads $\left<r^2\right>^{1/2} = 1.60b$. The parameter $\alpha = \left<u_1^2\right>/c_s^2$ is a measure of the temperature of the fluid, $T = M \left<u_1^2\right >$, where $M = \rho \dx^3$ is the mass of fluid in a single grid cell. The dimensionless time step in the FLB simulations is related to the temperature through the scaling with $t_0 = \xi/\kappa = \xi b^2/T$. $\Delta T$ is the difference in temperatures of the particles and fluid. The mass of a bead, kinematic viscosity of the fluid and the length of the periodic unit cell are $m = 0.1 M$, $\nu = 0.1 \dx^2/\dt$ and $L = 9.2R_g$, unless otherwise indicated. The statistical errors in diffusivity, $D/D_0$, and Rouse-mode relaxation time, $\tau_1/t_0$, are less than 0.5\%; $D_0 = T/\xi$ is the monomer diffusivity, and $Sc_0$ and $Sc$ are the Schmidt numbers based on the monomer and polymer diffusivities.}
\label{tab:props}
\footnotetext{$^1$ Length of periodic unit cell $L = 18.4R_g$. }
\footnotetext{$^2$ The viscosity of the fluid is varied: (a) $\nu = 0.05 \dx^2/\dt$;  (b) $\nu = 0.2 \dx^2/\dt$.}
\footnotetext{$^3$ The mass of the bead is varied: (a) $m = M$; (b) $m = 10M$. }
\footnotetext{ $^4$ Three-point interpolation.}
\footnotetext{ $^5$ No excitation of the kinetic (ghost) modes.}
\footnotetext{ $^6$ No excitation of the fluid modes.}
\footnotetext{ $^7$ Results from BD simulations.}
\end{table}

The fluctuation-dissipation theorem (FDT) should ensure that the polymer thermalizes to the same temperature as the fluid; in other words $m \left<v_i^2\right> = M \left<u_1^2\right>$, where $M = \rho \dx^3$ is the mass of fluid in a single LB grid cell. The column $\Delta T/T$ in Table~\ref{tab:props} measures the relative deviations in the polymer temperature from thermal equilibrium; these are usually small, of the order of $0.2\%$. Larger deviations occur when the temperature is too high ($\alpha = 0.024$), or when the LB model is not exactly thermalized (footnotes 5 and 6). If the kinetic (or ghost) modes are not subject to random forcing, the fluctuation-dissipation relation is broken at short length scales~\cite{Adhikari2005,Dunweg2007}. This causes a small error in the size of the polymer, $1-2\%$, when compared to the properly thermalized simulations, with similar deviations in the diffusion coefficient and Rouse relaxation times (footnote 5). Larger errors in both static and dynamic properties occur if the fluid dynamics is purely dissipative (footnote 6), because the fluctuation-dissipation relation is then broken at all length scales; results from simulations without fluid fluctuations, for example Ref.~\cite{Chen2007}, are invalid. The establishment of good thermal equilibrium between the polymer and fluid requires exact thermalization of the LB fluid and the coupling algorithm described in Sec.~\ref{sec:coupling}.

The ratios $\kappa/T$ and $\epsilon/T$ must be kept constant if the polymer conformations are to be independent of the degree of coarse graining of the fluid degrees of freedom. The dimensionless timestep $\dt/t_0$ of the FLB simulation in Fig.~\ref{fig:conf} is then controlled by the temperature of the fluctuating fluid  ($T$ or $\alpha$). Since the viscosity is independent of temperature in the FLB model, the Schmidt number $Sc = \eta / \rho D$, varies inversely with $T$. The results in Table~\ref{tab:props} show that the static and dynamic properties are both insensitive to Schmidt number, but there are small deviations when $Sc < 30$, which is consistent with earlier findings~\cite{Usta2005}. Other authors have suggested that the Schmidt number based on the monomer diffusion, $Sc_0 = \eta/\rho D_0$, should be in excess of 30~\cite{Ahlrichs1999}, but our results suggest that this may be overly restrictive; we do not find systematic deviations in either static or dynamic properties until $Sc_0 < 10$. Both static and dynamic properties are statistically independent of fluid viscosity (footnote 2) and particle mass (footnote 3) over the ranges studied. Finally, we note that the conformational properties and Rouse-mode relaxation times are independent of the size of the unit cell (footnote 1). These results are consistent with recent FLB simulations~\cite{Pham2009}, which found a weak system size dependence when $L < 5R_g$; in our simulations $L \sim 10R_g$. The systematic dependence of the diffusion coefficient on $L$ has been analytically corrected in Table~\ref{tab:props}, as discussed in detail in Sec.~\ref{sec:diff}. 

\subsection{Diffusion coefficient}\label{sec:diff}

The diffusion coefficient of the polymer is determined from the time-dependent displacement, $\Delta \Vr_c(t) = \Vr_c(t) - \Vr_c(0)$,  of the center of mass vector, $\Vr_c = (N+1)^{-1}\sum_{i=0}^N \Vr_i$. We calculate the diffusion coefficient from the time derivative,
\begin{equation}
D(t) = \frac{1}{6}\frac{d}{dt} \langle \Delta \Vr_c(t) \cdot \Delta \Vr_c(t) \rangle,
\label{eq:diff}
\end{equation}
rather than the slope, since the derivative asymptotes at much earlier times. The short-time diffusivity determined by Brownian dynamics is found from Eq.~\eqref{eq:diff} in the limit $t \rightarrow 0$. It is equal to the Kirkwood diffusivity and only slightly different, by $1 - 2\%$, from the long-time diffusivity~\cite{Liu2003}. The FLB simulations are inertial, and here $\lim_{t \rightarrow 0} D(t) = 0$. Nevertheless, in both methods the diffusivity reaches its asymptotic value, $D$, over a time of the order of the Zimm time, $t_Z = 6 \pi \eta R_g^3/T$.

\begin{figure}
\scalebox{0.5}{\includegraphics{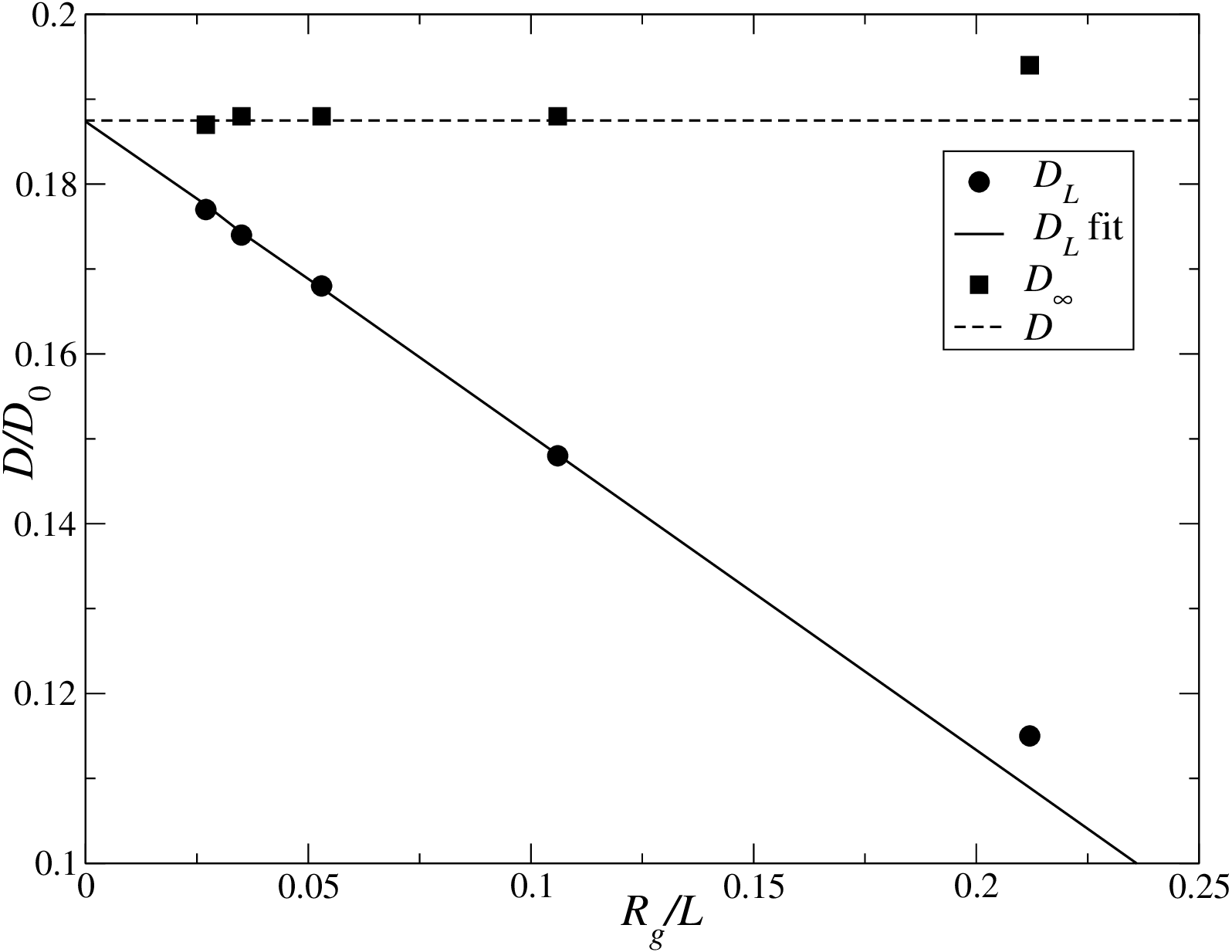}}
\caption{Diffusion coefficients from FLB simulations with different unit cell sizes, $L$. The coarsest grid resolution, $\dx/b = 2.58$, was used for computational efficiency.}
\label{fig:diff1}
\end{figure}

Our investigations show that the diffusion coefficient in an FLB simulation depends on only three parameters: the size of the periodic unit cell, $L$, the grid resolution, $\dx/b$, and the method of interpolation. Since the BD results are for an isolated polymer, the FLB data must be corrected for finite-size effects; here we use a well-established correction for the diffusion coefficient~\cite{Ladd1990a}, which effectively eliminates the dependence of the diffusion coefficient on the box size. The self-diffusion coefficient of a spherical particle of radius $R$ in a periodic system with repeat length $L$ can be written as~\cite{Ladd1990a},
\begin{equation}\label{eq:DL}
\frac{D_L}{D_\infty} = 1 - 2.837 \frac{R}{L} + \frac{4\pi R^3}{3L^3} + \ldots,
\end{equation} 
where $D_L$ is the diffusion constant in a system of length $L$ and $D_\infty = T/6 \pi \eta R$ is the diffusion coefficient of an isolated sphere.  At large distances, the average flow field around the polymer is similar to the flow field around a spherical particle; we therefore expect the same relation for the center-of-mass diffusion coefficient of the polymer.
Although we do not know {\em a priori} what the effective radius of the polymer is, the leading order correction is independent of $R$,
\begin{equation}\label{eq:Dcorr}
D_\infty = D_L + \frac{2.837 T}{6 \pi \eta L}.
\end{equation}

\begin{table}
\renewcommand*\arraystretch{0.65}
\begin{tabular}{|cD{.}{.}{6}|ccc|}
\hline
$L/R_g$ & \multicolumn{1}{c|}{$\alpha$} & $D_L/D_0$ & $D_\infty/D_0$ & $D_\infty^{(3)}/D_0$\\ \hline
2.8 & 0.0015 & 0.083 & 0.216 & 0.205\\
4.7 & 0.003 & 0.114 & 0.193 & 0.191 \\
4.7 & 0.0015 & 0.115 & 0.194 & 0.192\\
9.4 & 0.003 & 0.149 & 0.189 & 0.188\\
9.4 & 0.0015 & 0.148 & 0.188 & 0.187\\
18.9 & 0.003 & 0.168 & 0.188 & 0.188\\
18.9 & 0.0015 & 0.168 & 0.188 & 0.187\\
28.3 & 0.003 & 0.173 & 0.186 & 0.186\\
28.3 & 0.0015 & 0.170 & 0.184 & 0.184\\
28.3 & 0.00075 & 0.174 & 0.188 & 0.188\\
28.3 & 0.000375 & 0.174 & 0.188 & 0.188\\
37.7 & 0.003 & 0.173 & 0.183 & 0.183\\
37.7 & 0.0015 & 0.175 & 0.185 & 0.185\\
37.7 & 0.00075 & 0.176 & 0.186 & 0.186\\
37.7 & 0.000375 & 0.177 & 0.187 & 0.187\\ \hline
\end{tabular}
\caption{Effect of system size on the diffusion coefficient of a 10-segment chain. The resolution of the LB grid, $\dx/b = 2.58$. The parameter $\alpha = \left<u_1^2\right>/c_s^2$ is a measure of the temperature of the fluid, $T = M \left<u_1^2\right >$, where $M = \rho \dx^3$ is the mass of fluid in a single grid cell. $D_L$ is the diffusion coefficient from FLB simulations and $D_\infty$ is the corrected value from Eq~\eqref{eq:Dcorr}; results including the $(R/L)^3$ correction are indicated by $D_\infty^{(3)}$.}
\label{tab:diff}
\end{table}

The system-size dependence has been investigated using the coarsest resolution of the LB grid,  $\dx/b = 2.58$, to maximize computational efficiency. The results in Fig.~\ref{fig:diff1} show the expected linear dependence on $L^{-1}$, with the same asymptotic value of the polymer diffusivity ($D/D_0 = 0.1875$) from either extrapolation, fitting to 4 different system sizes ($9.5 < L/R_g < 38$), or from a single simulation with $L \sim 10R_g$. Since the computational cost scales as $L^3$, a single simulation takes $1/36$ the time of a sequence of three simulations with box lengths in the ratio $1:2:3$. Moreover, larger systems require a lower temperature to obtain the asymptotic (with $T$) diffusion coefficient, as shown in Table~\ref{tab:diff}. Data for large systems ($L > 20R_g$) shows that the limiting, low$-T$ diffusion coefficient, as plotted in Fig.~\ref{fig:diff1}, requires a temperature $4-8$ times smaller, which translates to $4-8$ times more processing for the same statistics.

It is possible to further refine the correction for finite-size effects by including the next term in Eq.~\eqref{eq:DL}, but this requires the polymer size. Defining the diffusivity of the isolated chain, $D_\infty = T/6\pi\eta R_\infty$, in terms of the effective radius $R_\infty$, and rearranging Eq.~\eqref{eq:DL} results in a cubic equation for $x = R_\infty/L$,
\begin{equation}\label{eq:DL3}
\frac{4}{3}\pi x^3 - (2.837+x_L)x + 1 = 0,
\end{equation}
where $x_L = R_L/L$, with $D_L= T/6\pi\eta R_L$. Since the additional correction is small, Eq.~\eqref{eq:DL3} can be solved for $x$ in a few iterations. This leads to slightly more consistent diffusion coefficients from the smaller box sizes, as shown in final column of Table~\ref{tab:diff}. The diffusivities in Table~\ref{tab:props} include the extra correction term.

\begin{figure}
\scalebox{0.5}{\includegraphics{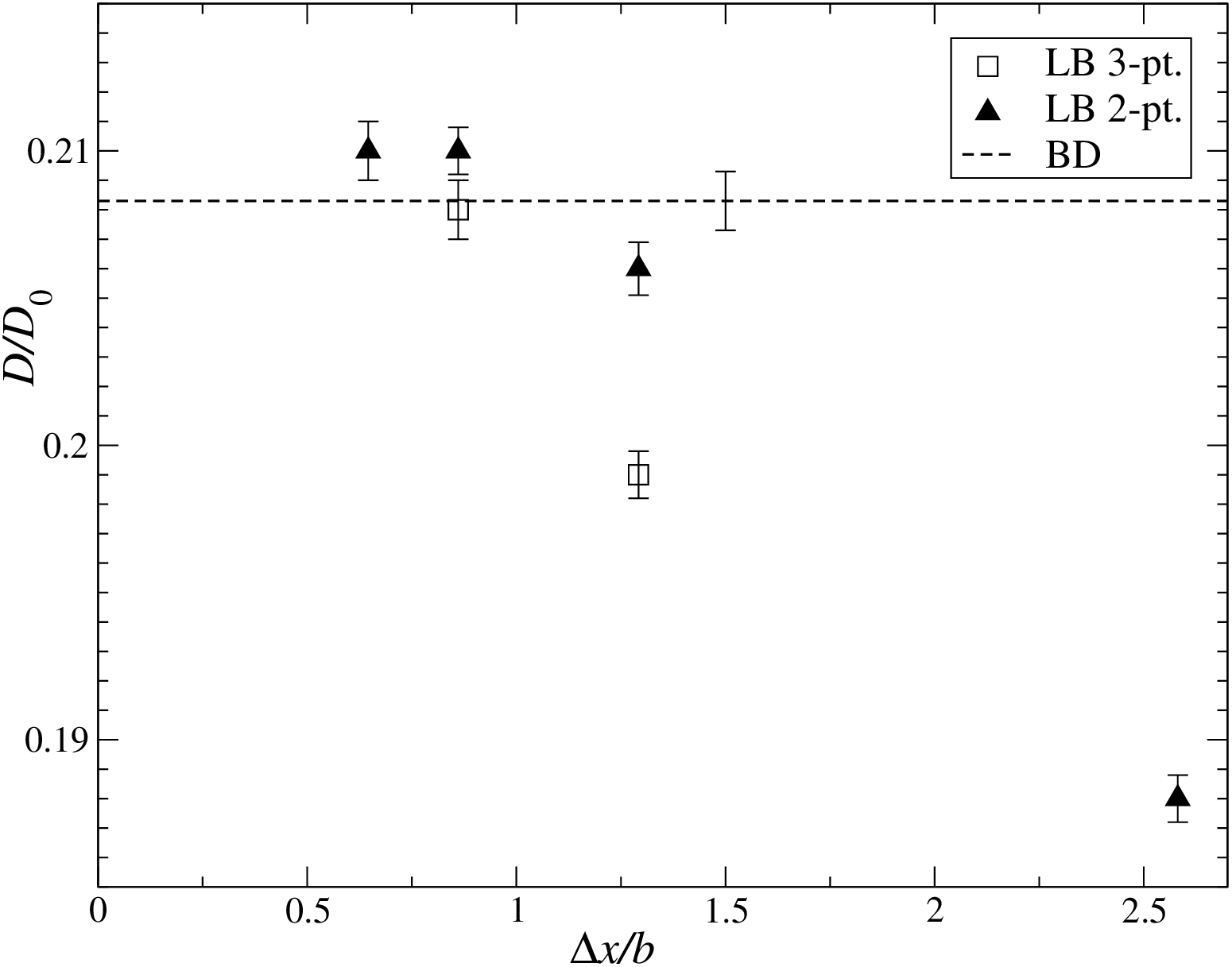}}
\caption{Diffusion coefficients from FLB simulations with different grid resolutions, $\dx/b$, are compared with Brownian dynamics (dashed line); the vertical bar indicates the statistical uncertainty in the BD data.}
\label{fig:diff2}
\end{figure}

Polymer diffusion coefficients from FLB simulations depend on grid resolution, $\dx$, and previous work~\cite{Ahlrichs1999,Usta2005} suggests that a ratio of $\dx/b \sim 1-2$ is adequate for most purposes. Here the diffusion coefficient of 10 segment chains are shown in Fig.~\ref{fig:diff2} for a range of different grid resolutions, $0.65 < \dx/b < 2.58$. The BD result is shown as a dashed line for clarity in comparison. By decreasing the ratio $\dx/b$, the number of grid points over which the typical hydrodynamic interactions are calculated is increased. The force coupling method is known to give an accurate representation of the hydrodynamic interactions when the distance between the beads is more than $3\dx$~\cite{Dunweg2009}, thus we expect to match the BD results for sufficiently small $\dx/b$.

For large grid-spacings, $\dx = 2.58b$, the FLB diffusivity is too small, in agreement with previous studies~\cite{Ahlrichs1999}, while for smaller grid spacings, $\dx < b$, an almost exact agreement between FLB and BD results is found. The effect of a large grid-spacing can be understood from the limiting case when the grid spacing exceeds the length of the entire chain, in which case the polymer dynamics follows the Rouse scaling~\cite{Ahlrichs1999}. The recommendation~\cite{Ahlrichs1999} that the grid spacing should be comparable to the mean distance between neighboring beads, $\dx \sim \left<r^2\right>^{1/2} = 1.60b$, leads to small errors in the diffusion constant, of the order of $2-3\%$.

\subsection{Rouse relaxation times}\label{sec:rouse}

\begin{figure}
\scalebox{0.5}{\includegraphics{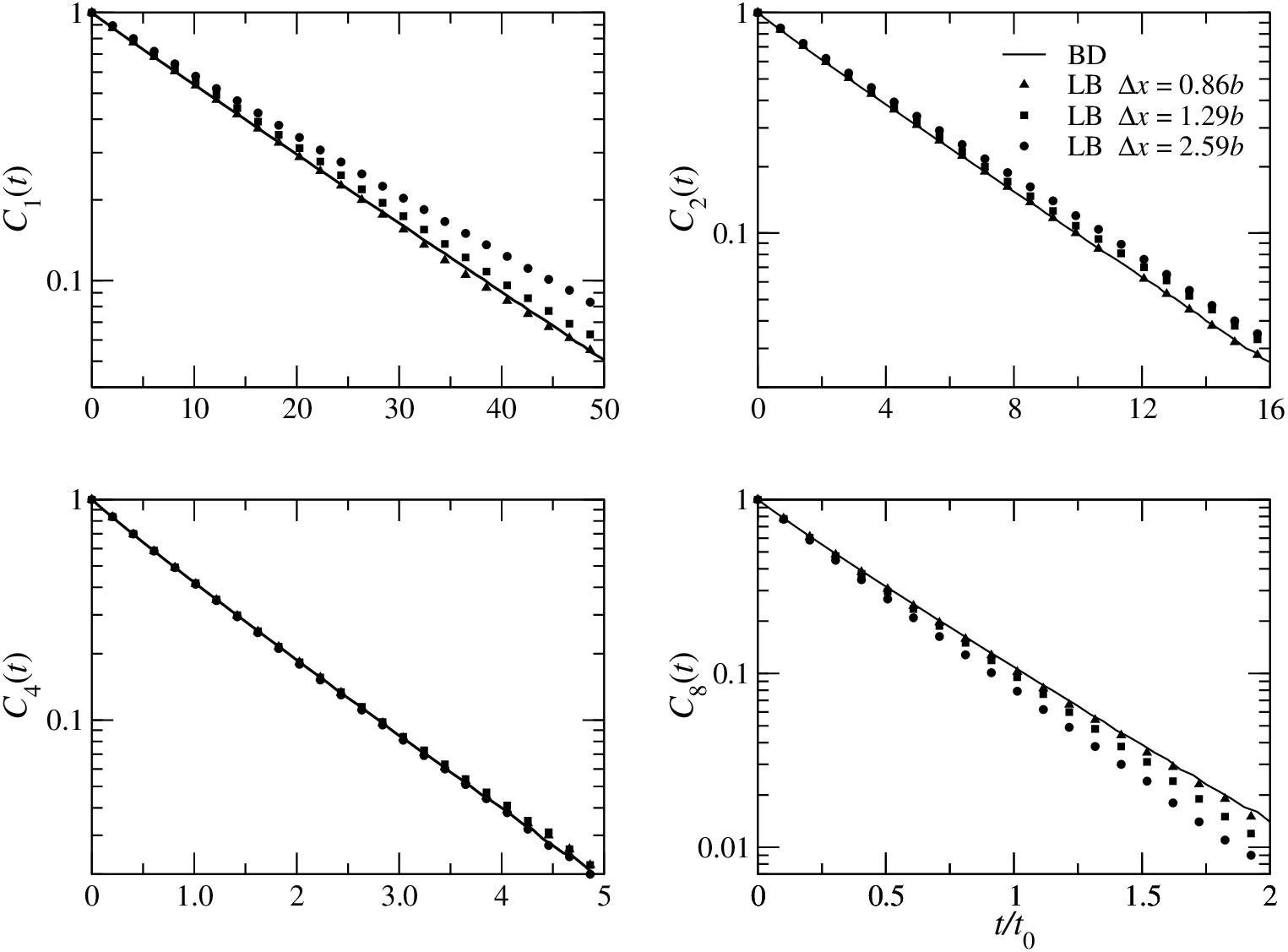}}
\caption{Normalized auto correlation functions of the Rouse-mode amplitudes, $C_p(t)$, Eq.~\eqref{eq:corr}. The two longest wavelength modes ($p = 1,2$), an intermediate wavelength ($p = 4$) and a short wavelength ($p = 8$) are shown for three different grid resolutions; $\dx/b = 2.58$ (circles), $\dx/b = 1.29$ (squares), and $\dx/b = 0.86$ (triangles); the solid lines are the Brownian dynamics results.}
\label{fig:rmcf}
\end{figure}

The internal motions of the polymer coil are a more sensitive measure of the hydrodynamic interactions than the center-of-mass diffusion coefficient. The polymer configuration can be represented by its normal (Rouse) modes $\mathbf{X}_p$~\cite{Doi1986},
\begin{equation}
\mathbf{X}_p = \frac{1}{N+1} \sum_{n=0}^{N} \mathbf{r}_n \cos \left[ \frac{p \pi}{N}(n+\frac{1}{2}) \right],  \quad p=1,2,.....\infty ,
\label{eq:xp}
\end{equation}
where $p$ denotes the mode number. The normalized autocorrelation function, $C_p(t)$, of the $p^{th}$ mode ($p > 0$),
\begin{equation}
C_p(t) = \frac{\langle \mathbf{X}_p(t)  \cdot \mathbf{X}_p(0) \rangle}{\langle \mathbf{X}_p(0) \cdot \mathbf{X}_p(0) \rangle},
\label{eq:corr}
\end{equation}
decays almost exponentially
\begin{equation}
C_p(t)  = \exp(-t / \mathbf{\tau}_p),
\label{eq:decay}
\end{equation}
as can be seen in Fig.~\ref{fig:rmcf}; $\mathbf{\tau}_p$ defines the Rouse relaxation time of the $p^{th}$ mode. The correlation functions for the two finest grid resolution, $\dx = 0.65b$ (not shown) and $\dx = 0.86b$, agree almost perfectly with Brownian dynamics, while for the coarsest resolution $\dx = 2.58b$ there are significant deviations in the long-wavelength modes. Somewhat surprisingly, the errors in the less resolved LB simulations diminish with increasing mode number, so that for $p = 4$ the results for all grid resolutions are essentially indistinguishable. However for still shorter wavelengths the discrepancies increase again for the coarser grids, this time in the opposite direction.

We have selected one particular time, $t = 12.6t_0$, where the $p= 1$ mode has decayed to $45\%$ of its initial value, for a more detailed comparison. For the three resolutions shown in Fig.~\ref{fig:rmcf}, the deviation from Brownian dynamics are $11\%$ ($\dx = 2.58b$), $5\%$ ($\dx = 1.29b$), and $< 0.5\%$ ($\dx = 0.86b$ and $\dx = 0.65b$), respectively. Data in Fig. 7 of Ref.~\cite{Pham2009}, taken at a similar time, show deviations between FLB and BD of the order of $2\%$. The grid resolution in these simulations corresponds to $\left<r^2\right>^{1/2} \approx 1.1 \dx$, similar to the intermediate resolution in our work, $\dx = 1.29b$ or $\left< r^2 \right>^{1/2} \approx 1.2 \dx$. Our results show slightly larger deviations, possibly due to the shorter chain or the softer excluded volume forces, both of which emphasize the short-range hydrodynamic interactions. For the system sizes we used, the ${\cal O} (L^{-3})$ corrections to the Rouse-mode relaxation times~\cite{Pham2009} are small; simulations with $\dx = 1.29b$ and $L = 18.8R_g$ (instead of $L = 9.4R_g$) show a similar ($4\%$) deviation from Brownian dynamics at $t = 12.6t_0$.

\begin{figure}
\scalebox{0.5}{\includegraphics{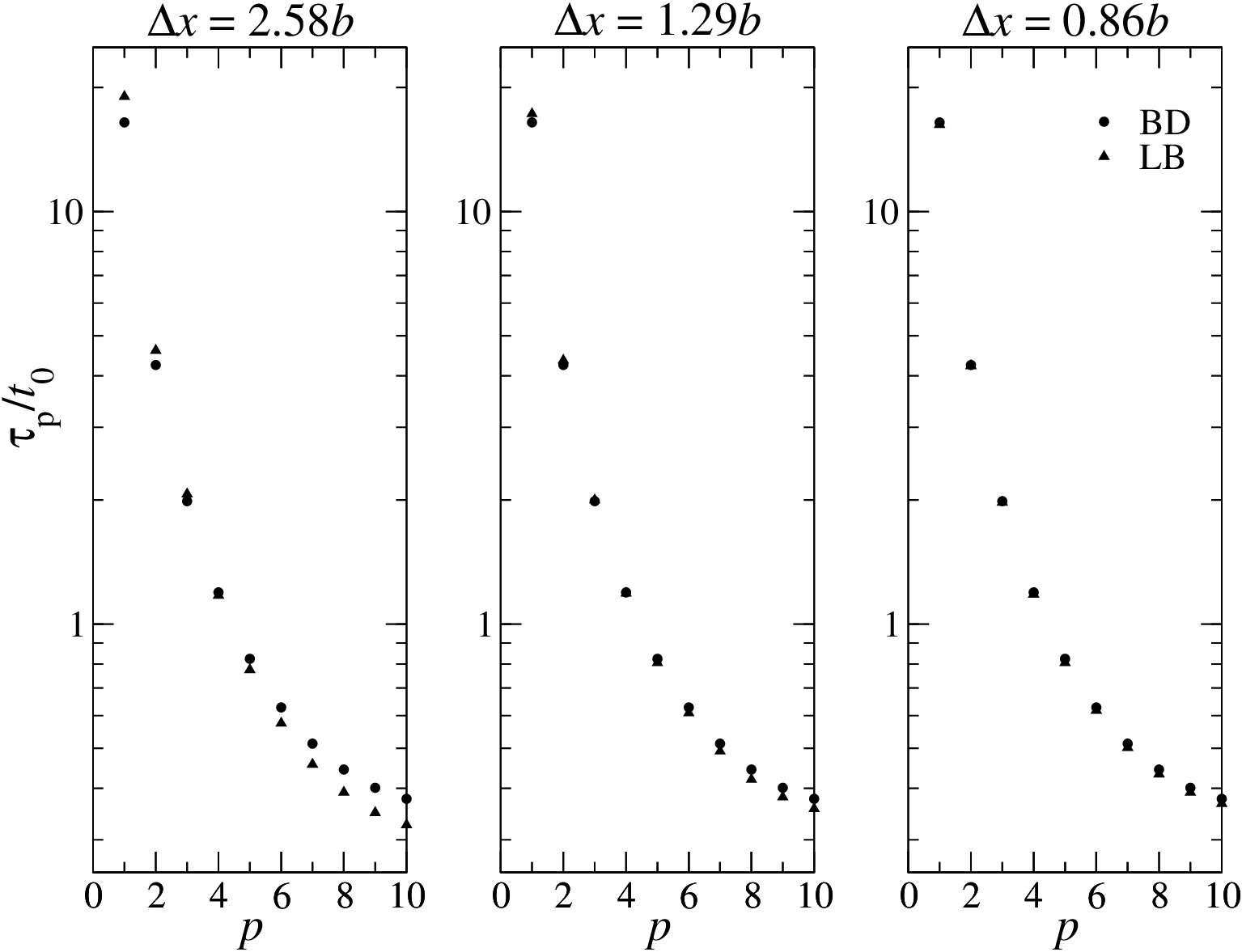}}
\caption{Relaxation times of the Rouse modes, $\tau_{p}$, from FLB (solid circles) and BD (solid triangles). Results  for $N = 10$ are compared for three different grid resolutions, $\dx = 2.58b$ (left), $\dx = 1.29b$ (center), and $\dx = 0.86b$ (right); results for $\dx = 0.65b$ are indistinguishable from $\dx = 0.86b$.}
\label{fig:rmrt}
\end{figure}

\begin{figure}
\scalebox{0.5}{\includegraphics{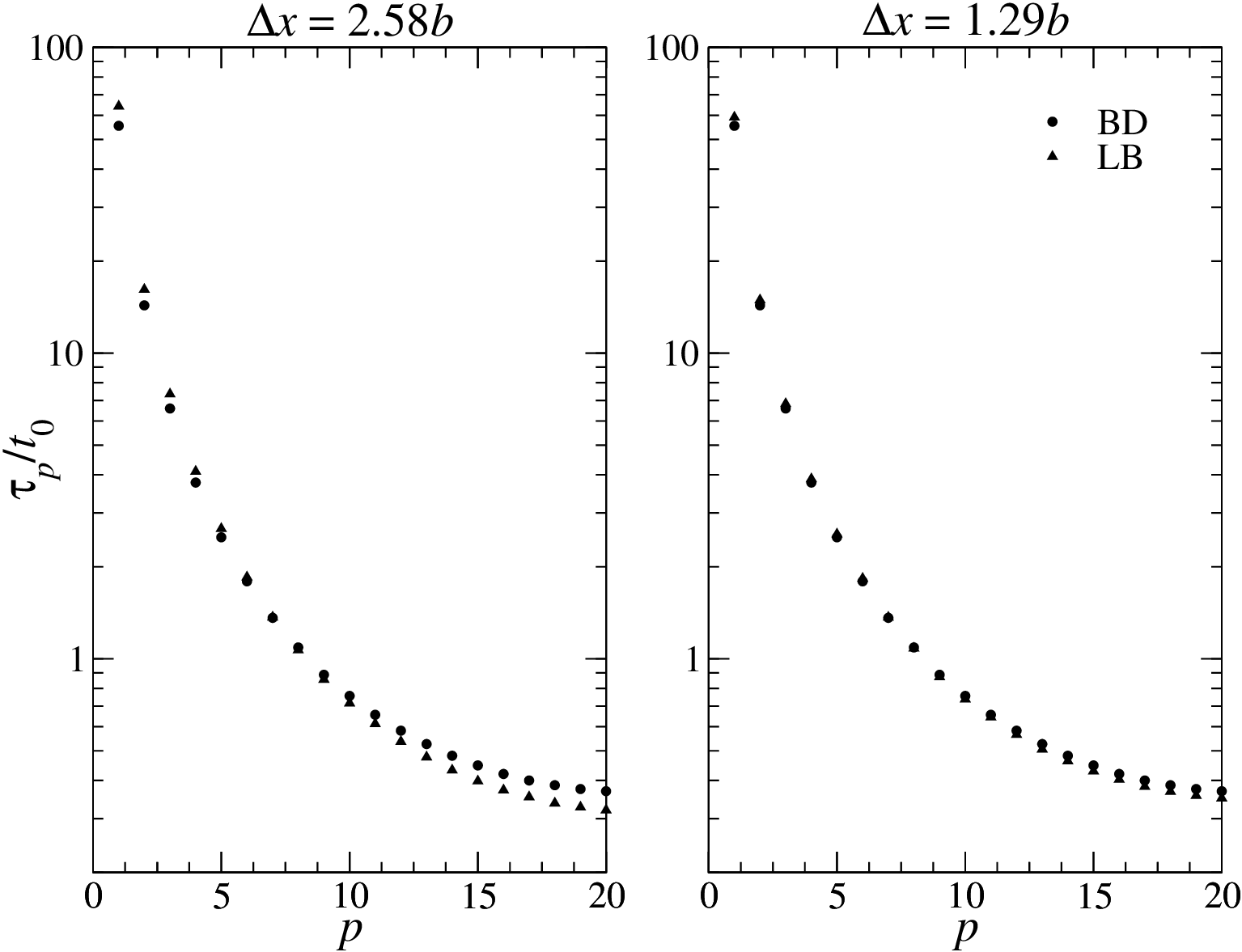}}
\caption{Relaxation times of the Rouse modes, $\tau_{p}$, from FLB (solid circles) and BD (solid triangles). Results  for $N = 20$ are compared for resolutions $\dx = 2.58b$ (left) and $\dx = 1.29b$ (right).}
\label{fig:rmrt20}
\end{figure}

The Rouse-mode relaxation times were calculated from the integral of the autocorrelation functions $C_p(t)$, Eq.~\eqref{eq:decay}. The first portion of the integral was calculated by numerical quadrature, up to a time where the correlation function had decayed to less than $5\%$ of its initial value. To minimize the effect of statistical errors on the integral, we fitted the last portion of the correlation function to a single exponential and calculated the long-time contribution analytically. The overall integral is insensitive to the exact value of the relaxation time of the fitted exponential, which was determined self-consistently from the value of $\tau_p$. Since the decay of $C_p(t)$ follows a single exponential almost exactly, this procedure is quite precise; we used the same protocol for both FLB and BD correlation functions.

The data in Fig.~\ref{fig:rmrt} show that the lattice-Boltzmann method can reproduce the whole Rouse spectrum when sufficiently resolved. For the two finest resolutions, $\dx = 0.86b$ and $\dx = 0.65b$ (not shown), the deviations in the long-wavelength relaxation times are less than $1\%$, and are only slightly larger ($\sim 2\%$) at short wavelengths. At the intermediate resolution, $\dx = 1.29b$, the long-wavelength relaxation times are well represented, with errors of $5\%$ at most, but at the coarsest resolution the deviations are larger, up to $15\%$. Finally, in Fig.~\ref{fig:rmrt20} we compare the Rouse spectrum for chains with 20 segments. The errors in the relaxation times are similar to those obtained for the shorter chains with the same grid resolution.

\section{Conclusions}\label{conclusions}

This comparative study of lattice-Boltzmann and Brownian dynamics simulations of a semi-flexible polymer demonstrates that static and dynamic properties of isolated chains agree quantitatively, within $1-2\%$. Our paper complements recently published work~\cite{Pham2009} through a systematic investigation of variations in the LB model parameters. Our results show that hydrodynamic retardation, which is sometimes suggested to be a reason for discrepancies between LB and BD results~\cite{Chen2007}, is in fact easily controlled; the diffusion coefficient and Rouse spectrum are independent of Schmidt number when $Sc_0 > 10$. Other parameters such as fluid viscosity and bead mass have little effect on the results. Somwhat disappointingly, a higher-order interpolation of the fluid velocity field does not lead to improved agreement with Brownian dynamics. Although results with three-point interpolation converge to the same values as with linear interpolation, the convergence is slower, rather than faster as one would have hoped. Despite the smoother interpolation of the flow field, the force is delocalized over a larger volume and this seems to reduce the accuracy, while simultaneously increasing the computational cost.

The crucial parameter affecting the accuracy of a lattice-Boltzmann simulation is the resolution of the polymer on the LB grid. When the mean distance between neighboring beads is more than twice the LB grid spacing, the agreement between FLB and BD simulations is essentially exact. However the computational cost of a fine grid is high, scaling as the resolution to the sixth power. Thus a reasonable practical compromise is the original suggestion $\left< r^2 \right>^{1/2} \approx \dx$~\cite{Ahlrichs1999}. The errors in the dynamic properties are then around $5\%$, which is sufficient for most purposes. A typical LB simulation, $\dx = 1.29b$, $L = 9.4R_g$, $\alpha = 0.003$, run for $8 \times 10^5 t_0$, requires about 70 hours of computation. Comparable Brownian dynamics simulations take approximately one hour. However, simulations of concentrated solutions in confined geometries are more favorable for lattice-Boltzmann methods~\cite{Pham2009}.

\bibliography{suspensions,polymers,lg_lb,md,misc,./pre_09}
\end{document}

%% file: macros.tex
\newcommand{\BB}[1]		{\mbox{\boldmath${#1}$}}
\newcommand{\modulo}[1]	{{\left \vert {#1} \right \vert}}

\newcommand{\Vc}		{{\BB{c}}}

\newcommand{\Ve}		{{\BB{e}}}

\newcommand{\Vj}		{{\BB{j}}}

\newcommand{\Vn}		{{\BB{n}}}

\newcommand{\Vp}		{{\BB{p}}}

\newcommand{\Vr}		{{\BB{r}}}

\newcommand{\Vu}		{{\BB{u}}}
\newcommand{\Vv}		{{\BB{v}}}
\newcommand{\Vw}		{{\BB{w}}}

\newcommand{\VF}		{{\BB{F}}}
\newcommand{\VI}		{{\BB{I}}}

\newcommand{\VR}		{{\BB{R}}}

\newcommand{\Vone}		{{\BB{1}}}

\newcommand{\Vmu}		{{\BB{\mu}}}

\newcommand{\Vphi}		{{\BB{\phi}}}

\newcommand{\Dx}		{\BB{\nabla}}

\newcommand{\cf}		{{\it c.f. }}

\newcommand{\dx}		{\Delta x}
\newcommand{\dt}		{\Delta t}